%PAGE SIZE, FONTS, PREPRINT NUMBER ETC.
\magnification=1200\overfullrule=0pt\baselineskip=14.5pt
\vsize=23truecm \hsize=15.5truecm \overfullrule=0pt\pageno=0

\font\titlefont=cmbx10 scaled \magstep1
\font\sectnfont=cmbx8  scaled \magstep1
\def\mname{\ifcase\month\or January \or February \or March \or April
           \or May \or June \or July \or August \or September
           \or October \or November \or December \fi}
\def\date{\hbox{\strut\mname \number\year}}
\def\crnum{\hbox{HLRZ 65/93 \strut}}
\def\banner{\hfill\hbox{\vbox{\crnum}}\relax}
\def\manner{\hbox{\vbox{\offinterlineskip\bigskip\bigskip
                        \crnum\date}}\hfill\relax}
\footline={\ifnum\pageno=0\manner\else\hfil\number\pageno\hfil\fi}
%
%
% Numbering of figures, tables and equations is automatic.
%
\newcount\FIGURENUMBER\FIGURENUMBER=0
\def\figtag#1{\expandafter\ifx\csname FG#1\endcsname\relax
               \global\advance\FIGURENUMBER by 1
               \expandafter\xdef\csname FG#1\endcsname
                              {\the\FIGURENUMBER}\fi
              \csname FG#1\endcsname\relax}
\def\fig#1{\expandafter\ifx\csname FG#1\endcsname\relax
               \global\advance\FIGURENUMBER by 1
               \expandafter\xdef\csname FG#1\endcsname
                      {\the\FIGURENUMBER}\fi
           Fig.~\csname FG#1\endcsname\relax}
\newcount\TABLENUMBER\TABLENUMBER=0
\def\tabletag#1{\expandafter\ifx\csname TB#1\endcsname\relax
               \global\advance\TABLENUMBER by 1
               \expandafter\xdef\csname TB#1\endcsname
                          {\the\TABLENUMBER}\fi
             \csname TB#1\endcsname\relax}
\def\table#1{\expandafter\ifx\csname TB#1\endcsname\relax
               \global\advance\TABLENUMBER by 1
               \expandafter\xdef\csname TB#1\endcsname{\the\TABLENUMBER}\fi
             Table \csname TB#1\endcsname\relax}
\newcount\REFERENCENUMBER\REFERENCENUMBER=0
\def\reftag#1{\expandafter\ifx\csname RF#1\endcsname\relax
               \global\advance\REFERENCENUMBER by 1
               \expandafter\xdef\csname RF#1\endcsname
                      {\the\REFERENCENUMBER}\fi
             \csname RF#1\endcsname\relax}
\def\ref#1{\expandafter\ifx\csname RF#1\endcsname\relax
               \global\advance\REFERENCENUMBER by 1
               \expandafter\xdef\csname RF#1\endcsname
                      {\the\REFERENCENUMBER}\fi
             [\csname RF#1\endcsname]\relax}
\def\refto#1#2{\expandafter\ifx\csname RF#1\endcsname\relax
               \global\advance\REFERENCENUMBER by 1
               \expandafter\xdef\csname RF#1\endcsname
                      {\the\REFERENCENUMBER}\fi
           \expandafter\ifx\csname RF#2\endcsname\relax
               \global\advance\REFERENCENUMBER by 1
               \expandafter\xdef\csname RF#2\endcsname
                      {\the\REFERENCENUMBER}\fi
             [\csname RF#1\endcsname--\csname RF#2\endcsname]\relax}
\def\refs#1#2{\expandafter\ifx\csname RF#1\endcsname\relax
               \global\advance\REFERENCENUMBER by 1
               \expandafter\xdef\csname RF#1\endcsname
                      {\the\REFERENCENUMBER}\fi
           \expandafter\ifx\csname RF#2\endcsname\relax
               \global\advance\REFERENCENUMBER by 1
               \expandafter\xdef\csname RF#2\endcsname
                      {\the\REFERENCENUMBER}\fi
            [\csname RF#1\endcsname,\csname RF#2\endcsname]\relax}
\newcount\EQUATIONNUMBER\EQUATIONNUMBER=0
\def\EQNO#1{\expandafter\ifx\csname EQ#1\endcsname\relax
               \global\advance\EQUATIONNUMBER by 1
               \expandafter\xdef\csname EQ#1\endcsname
                      {\the\EQUATIONNUMBER}\fi
            \eqno(\csname EQ#1\endcsname)\relax}
\def\eq#1{\expandafter\ifx\csname EQ#1\endcsname\relax
               \global\advance\EQUATIONNUMBER by 1
               \expandafter\xdef\csname EQ#1\endcsname
                      {\the\EQUATIONNUMBER}\fi
          Eq.~(\csname EQ#1\endcsname)\relax}
\def\eqand#1#2{\expandafter\ifx\csname EQ#1\endcsname\relax
               \global\advance\EQUATIONNUMBER by 1
               \expandafter\xdef\csname EQ#1\endcsname
                        {\the\EQUATIONNUMBER}\fi
          \expandafter\ifx\csname EQ#2\endcsname\relax
               \global\advance\EQUATIONNUMBER by 1
               \expandafter\xdef\csname EQ#2\endcsname
                      {\the\EQUATIONNUMBER}\fi
         Eqs.~(\csname EQ#1\endcsname) and (\csname EQ#2\endcsname)\relax}
%
%DEFINE JOURNAL NAMES
        
    \def\JSP{{\sl J.\ Stat.\ Phys.\ }}
\def\NP{{\sl Nucl.\ Phys.\ }}         \def\PL{{\sl Phys.\ Lett.\ }}
\def\PR{{\sl Phys.\ Rev.\ }}          
\def\PRL{{\sl Phys.\ Rev.\ Lett.\ }}  
\def\ZP{{\sl Z.\ Phys.\ }}
\def\T{{\scriptscriptstyle T}} \def\V{{\scriptscriptstyle V}}
\def\CD{\xi_d}\def\cd{\ifmmode\CD\else$\CD$\fi}
\def\CO{\xi_o}\def\co{\ifmmode\CO\else$\CO$\fi}
\def\SOO{\sigma_{oo}}\def\soo{\ifmmode\SOO\else$\SOO$\fi}
\def\SOD{\sigma_{od}}\def\sod{\ifmmode\SOD\else$\SOD$\fi}
\def\lb{\hfil\penalty-10000}
 \def\ie{{\sl i.e.\/}} \def\etal{{\sl et al.\/}}
\newdimen\digitwidth\setbox0=\hbox{\rm0}\digitwidth=\wd0
{\vsize=19truecm\hsize=14truecm\voffset=1truecm\hoffset=0.5truecm
\banner\bigskip\bigskip\bigskip\baselineskip=15pt
\begingroup\titlefont\obeylines
\hfil INSTANTONS AND SURFACE TENSION\hfil
\hfil AT A FIRST-ORDER TRANSITION\hfil
\endgroup\bigskip
\bigskip\centerline{Sourendu Gupta}
\centerline{HLRZ, c/o KFA J\"ulich, D-5170 J\"ulich, Germany.}
\vskip4truecm
\centerline{\bf ABSTRACT}\medskip
We study the dynamics of the first order phase transition in the two
dimensional 15-state Potts model, both at and off equilibrium. We find
that phase changes take place through nucleation in both cases, and
finite volume effects are described well through an instanton
computation. Thus a dynamical measurement of the surface tension is
possible. We find that the order-disorder surface tension is compatible
with perfect wetting. An accurate treatment of fluctuations about the
instanton solution is seen to be of great importance.
\vfil\eject}

First order phase transitions, \ie, phase coexistence points, have
recently been subjected to intense analysis. Such systems are
characterised by many different dimensionful quantities. It is useful
to divide these into two classes of observables. The first pertains to
properties of the pure phases. Such properties are obtained,
as usual, by taking derivatives of the (extensive) free energy. These
derivatives, or cumulants, are extracted through finite-size scaling.
In recent years a full theory of such scaling has been developed
\refs{borgs}{fss} and tested \refto{borgs}{billo}.
The second class concerns coexistence.
Foremost among such variables is the surface
tension--- the leading non-extensive part of the free energy. Most
theories of the (canonical) dynamics at the phase transition involve the
surface tension. The measurement of this quantity is usually approached
through detailed investigations of the static system at the phase
transition \ref{sten}.

However, much of the recent attention enjoyed by first-order phase
transitions is due to the interesting dynamics of the transition. This
is expected to be due to nucleation. We perform a careful analysis of
finite size effects in the equilibrium dynamics and compare our
observations with an instanton-based computation \ref{zinn}. We
also test nucleation theory through a non-equilibrium process---
hysteresis. Certain scaling laws for this have been proposed recently
\ref{deepak}; we verify them for the first time. These two tests provide
the justification for the use of dynamical techniques for the
measurement of the surface tension.

Our numerical work is done with a simple model--- the two dimensional 15
state Potts model. This model is defined through the partition function
and Hamiltonian
$$ Z\;=\;\sum_{\{\sigma_r\}}\exp(-\beta H),\qquad
     H\;=\;\sum_{\langle rr'\rangle} 1-\delta_{\sigma_r\sigma_r'},
   \EQNO{model}$$
where the spin $\sigma_r$ sitting at the site $r$ of a (square) lattice
can take one of 15 values and the angular brackets denote nearest
neighbours. At $\beta_t=\log(\sqrt{15}+1)$, ordered and disordered
phases coexist in the thermodynamic limit. The two phases can be
distinguished either through a singlet magnetisation or the internal
energy density
$$ e\;=\; {1\over L^2}\langle H\rangle, \EQNO{eden}$$
where the lattice size is $L^2$. In the thermodynamic limit, the internal
energies in the ordered and disorder phases are, respectively, $e_0=1.779$
and $e_d=0.737$.

Recent analysis of correlation lengths in Potts models \ref{bw} have
yielded an exact result for a spin-spin correlation length. This has
been identified with the correlation length in the disordered phase
\cd. A duality argument has been used \ref{bj} to relate the order-order
surface tension, \soo, to this correlation length---
$$\soo\;=\;1/\cd.     \EQNO{sodef}$$
The perfect wetting conjecture would then imply the relation
$$2\sod\;=\;\soo\;=\;1/\cd.   \EQNO{pw}$$
between the order-disorder surface tension, \sod, and the rest of these
quantities. Although proven only in the limit $q\to\infty$, it is
strongly suspected that perfect wetting holds for two dimensional Potts
models for all $q>4$. From the formul\ae{} of \ref{bw} one finds $\cd=
4.18$ for $q=15$; implying $2\sod=0.239$. Thus a quantitative test of
nucleation theory should yield this value for \sod. Alternately, the
argument can be turned around and the measurement of the single quantity,
\sod, can be used to check the perfect wetting conjecture.

One of the dynamical methods we use is a finite-size scaling of the
exponential autocorrelation time, $\tau$, of the energy density.
Through a study of the 10-state Potts model, numerical evidence
was presented in \ref{us} that this autocorrelation time is determined by
the tunnelling rate between the coexisting ordered and disordered phases.
As a result, the autocorrelation time can be identified with the outcome
of an instanton computation \ref{zinn} giving
$$\tau^{-1}(L)\;=\; aL^{-d/2}\exp(-2\sod L^{d-1}).   \EQNO{zinn}$$
In our case, of course, $d=2$. The exponential factor is the
saddle-point result; and the $L$-dependence of the prefactor is
obtained from a one-loop computation of the determinant of the
fluctuations around the saddle point.
The test of the instanton computation is in the $L$-dependence of the
measured values of $\tau$. Note that \eq{zinn} describes dynamics in
equilibrium.

We also study a particular example of off-equilibrium dynamics, that of
hysteresis. The coupling $\beta$ is cyclically varied about the critical
coupling $\beta_t$ with a frequency $\omega$ and an amplitude $\Delta
\beta$. Hysteresis occurs as $e$ switches between the values $e_o$ and
$e_d$.  The area of the hysteresis loop in the energy density,
$A(\omega)$ is studied as a function of $\omega$. The system has a
real-time excitation with a `mass' given by \eq{zinn}. The measurement
of $A(\omega)$ is really a `line-shape' measurement. When $\Delta\beta
\to0$, the peak should be at $\tau^{-1}(L)$ and the shape should be
approximately Lorentzian. This cannot be converted to an useful test
because the functional form of the finite $\Delta\beta$ corrections is
not known.

We define the coercive coupling $\delta\beta_c$ through the fact that
hysteresis loops reach the value $(e_o+e_d)/2$ when the coupling
is $\beta_t\pm\delta\beta_c$. Here the nucleation rate becomes larger
than $1/\omega$, and the probability of a flip into the stable phase
exceeds $1/2$. This argument was presented in \ref{deepak} and was
developed into the scaling law
$$A(\omega)\sim (\ln\omega)^{-1/(d-1)}\qquad{\rm where\ }\omega\to0.
      \EQNO{deepak}$$
The coercive coupling also obeys the same scaling law.
Verification of this relation thus furnishes a test of nucleation
theory. Furthermore, since the tunnelling rate is finite for any finite
lattice, $A(\omega)$ is zero at some non-zero frequency $\Omega$, and we
have the relation
$$\Omega(L)\;=\;\tau^{-1}(L)\qquad{\rm where\ }A(\Omega)=0.
    \EQNO{omg}$$
Thus the scaling of $\Omega$ with $L$ is again given by \eq{zinn}, and
constitutes yet another test of nucleation theory. Note that \eq{deepak}
refers to a slow non-equilibrium situation.

\midinsert\centerline{TABLE \tabletag{runs}.}\medskip
Run parameters for the two-dimensional 15-state Potts model. We show the
lattice sizes $L$, pseudo-critical couplings $\beta_\T(L)$, statistics
used for the determination of autocorrelation times, $N_{run}$, and the
hysteresis parameters $\beta_m$, $\Delta\beta$ and $\delta\beta$.
{\catcode`?=\active\def?{\kern\digitwidth}
$$\vbox{\offinterlineskip\halign{
      \vrule#&\strut\hfil$\;#\;$\hfil&\vrule#&\hfil$\;#\;$\hfil&
      \vrule#&\hfil$\;#\;$\hfil&\vrule#&\hfil$\;#\;$\hfil&
      \vrule#&\hfil$\;#\;$\hfil&\vrule#&\hfil$\;#\;$\hfil&
      \vrule#&\hfil$\;#\;$\hfil&\vrule#&\hfil$\;#\;$\hfil&
      \vrule#&\hfil$\;#\;$\hfil&\vrule#&\hfil$\;#\;$\hfil&
      \vrule#&\hfil$\;#\;$\hfil&\vrule#&\hfil$\;#\;$\hfil&\vrule#\cr
\noalign{\hrule}
&L&&\beta_\T(L)&&N_{run}&&\multispan5\hfil{set 1}\hfil
                        &&\multispan5\hfil{set 2}\hfil
                        &&\multispan5\hfil{set 3}\hfil&\cr
&\omit&& && &&\multispan{17}\hrulefill&\cr
& && &&           &&\beta_m&&\Delta\beta&&\delta\beta
                  &&\beta_m&&\Delta\beta&&\delta\beta
                  &&\beta_m&&\Delta\beta&&\delta\beta&\cr
\noalign{\hrule}
&?8&&1.546?&&1\times10^6&&1.65  &&0.200 &&0.010
                        &&      &&      &&
                        &&      &&      &&      &\cr
&12&&1.5661&&1\times10^6&&1.65  &&0.100 &&0.005
                        &&1.61  &&0.060 &&0.005
                        &&1.6470&&0.1620&&0.0135&\cr
&16&&1.5723&&2\times10^6&&1.61  &&0.080 &&0.005
                        &&1.61  &&0.060 &&0.005
                        &&1.6260&&0.108?&&0.009?&\cr
&20&&1.5772&&4\times10^6&&1.61  &&0.060 &&0.005
                        &&      &&      &&
                        &&      &&      &&      &\cr
&24&&1.5798&&5\times10^6&&      &&      &&
                        &&      &&      &&
                        &&      &&      &&      &\cr
&30&&1.5811&&4\times10^7&&      &&      &&
                        &&      &&      &&
                        &&      &&      &&      &\cr
\noalign{\hrule}}}$$}\endinsert

For Potts models at phase coexistence it was observed \ref{us} that
both local and Swendson-Wang dynamics are dominated by tunnellings,
and that the exponential autcorrelation times, with changing $\beta$
and $L$, are related by a constant. In view of this, all our simulations
were performed with the latter algorithm. Autocorrelation times were
measured at the pseudo-critical couplings, $\beta_\T(L)$, defined by
the maximum of the specific heat. For $L\le20$ the values of
$\beta_\T(L)$ and $\tau$ were obtained in \ref{matthias}; this work
verifies these measurements.
Hysteresis was induced by cyclically changing $\beta$ from a maximum
of $\beta_m$ down by an amount $\Delta\beta$ and back, in discrete
steps of $\delta\beta$, running $N$ sweeps of a Swendson-Wang update
at each coupling. The runs were started by first thermalising a system
at $\beta_m$ with $2\times10^5$ cluster updates. Then for each $N$ (and
fixed values of the other parameters) we ran through 200 hysteresis
cycles. Since fairly large values of $N$ had to be used, this was by far
the most CPU-intensive part of these computations. The run parameters are
shown in \table{runs}. Three different sets of parameters were used in
order to check that the scaling law of \eq{deepak} and the extrapolated
values of $\Omega(L)$ were independent of $\Delta\beta$.

For each hysteresis loop, we defined the area by the sum
$$A\;=\; \delta\beta\sum_i (-1)^D \overline{e_i}, \EQNO{numarea}$$
where $i$ labels each of the values of $\beta$ in the cycle, the bar above
$e_i$ denotes averaging over the $N$ sweeps performed at that $\beta$. The
value of $D$ was set to be equal to 0 in that half of the cycle with
decreasing $\beta$, and 1 in the other. The averages and errors
were obtained by jack-knife estimators over all the cycles.

A second measurement was of the cyclic response function
$$c_\V^i\;=\; \langle (e_i-\langle e_i\rangle)^2\rangle,   \EQNO{resp}$$
where the angular brackets denote averages over all measurements performed
at a given coupling in the hysteresis cycle.
This response function peaks twice during a cycle, at
$\pm\delta\beta_c$, and allows us to extract $\delta\beta_c$ by
searching for the maximum of $c_\V^i$. This procedure is the dynamical
analogue of defining the transition coupling on a finite lattice,
$\beta_\T(L)$ by the peak in usual response function $c_\V$.

The average and error were again estimated by a jack-knife procedure.
These measurements of $\delta\beta_c$ have larger relative errors
than $A$. This is due to two reasons. The first
is intrinsic. Since tunnellings occur at random times, there is a cycle to
cycle variation in the coupling at which tunnellings occur. The other error
is related to the statistics. The identification of the coercive coupling
depends on the relative heights of the two peaks, and is subject to
fairly strong errors. Due to these uncertainties, we decided not to
use this quantity for our scaling tests.

The frequency $\omega$ should be identified with
$2\pi\delta\beta/N\Delta\beta$.
However, for each $L$, since $\delta\beta$ as well as
$\Delta\beta$ are fixed inside each set, these factors are not
important when trying to check the scaling with $\omega$. Furthermore,
when comparing different values of $L$, we are interested
in time scales expressed directly in sweeps. Hence we shall use the
convention $\omega=1/N$. This is only a matter of convenience.
When necessary, one should use the full definition of $\omega$.

\midinsert\vskip9truecm
\centerline{FIGURE \figtag{lorentz}}
\includegraphics{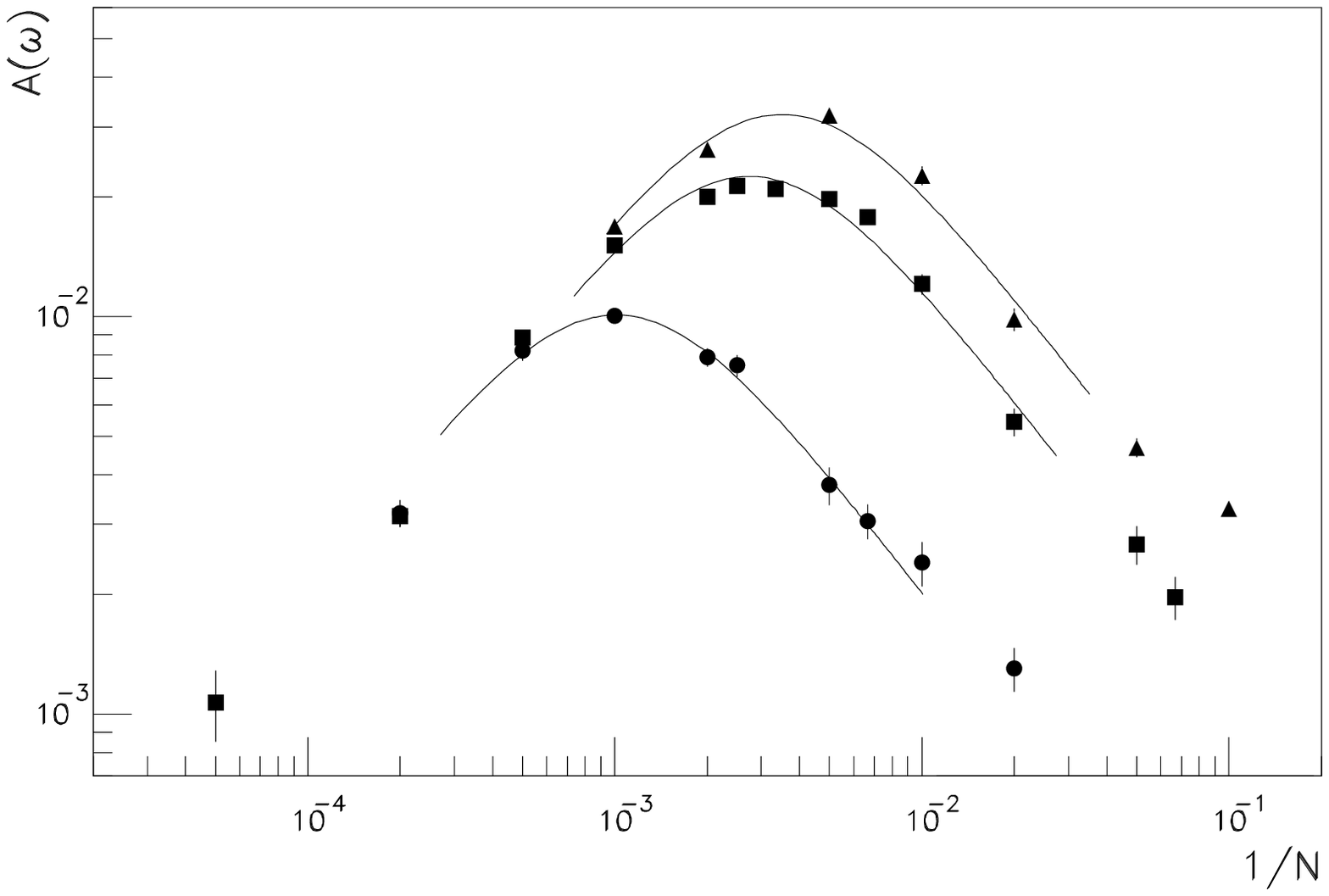}
Hysteresis loop areas $A(\omega)$ as functions of $1/N$ for the
15-state Potts model on $16^2$ lattices. The data are for sets 1 (squares),
2 (circles) and 3 (triangles) of \table{runs}. The lines show the
best Lorentzian fits.
\endinsert

For each $L$ and a set of $\omega$ at fixed $\Delta\beta$ and
$\delta\beta$, we tried to fit the data on $A(\omega,\Delta\beta)$ to a
Lorentzian. We found that this description improves as $\Delta\beta$
decreases. This is illustrated in \fig{lorentz} for the $L=16$ lattice.
At low frequencies, where the data deviates from the Lorentzian shape,
$A(\omega)$ is independent of $\Delta\beta$.
For large $N$, when the data deviate from the Lorentzian shape,
we fit a form
$$A(\omega)\;=\; a - b/\log N.    \EQNO{fitform}$$
We found extremely good fits to this form and could certainly rule out
any power-law behaviour. The data and fits are shown in \fig{nucl}.
The fitted parameters $a$ and $b$ give estimates of $\Omega$---
the frequency at which the loop area vanishes. Errors on $\Omega$
were estimated from the covariance matrix between these parameters.
We found good agreement between these values obtained indirectly and
the direct measurements of $\tau(L)$. The scaling of $\Omega^{-1}(L)$
with $L$ is consistent with \eq{zinn}, but does not provide a very
stringent test. The direct measurements of $\tau(L)$ are, of course,
more accurate.

\midinsert\vskip9truecm
\centerline{FIGURE \figtag{nucl}}
\includegraphics{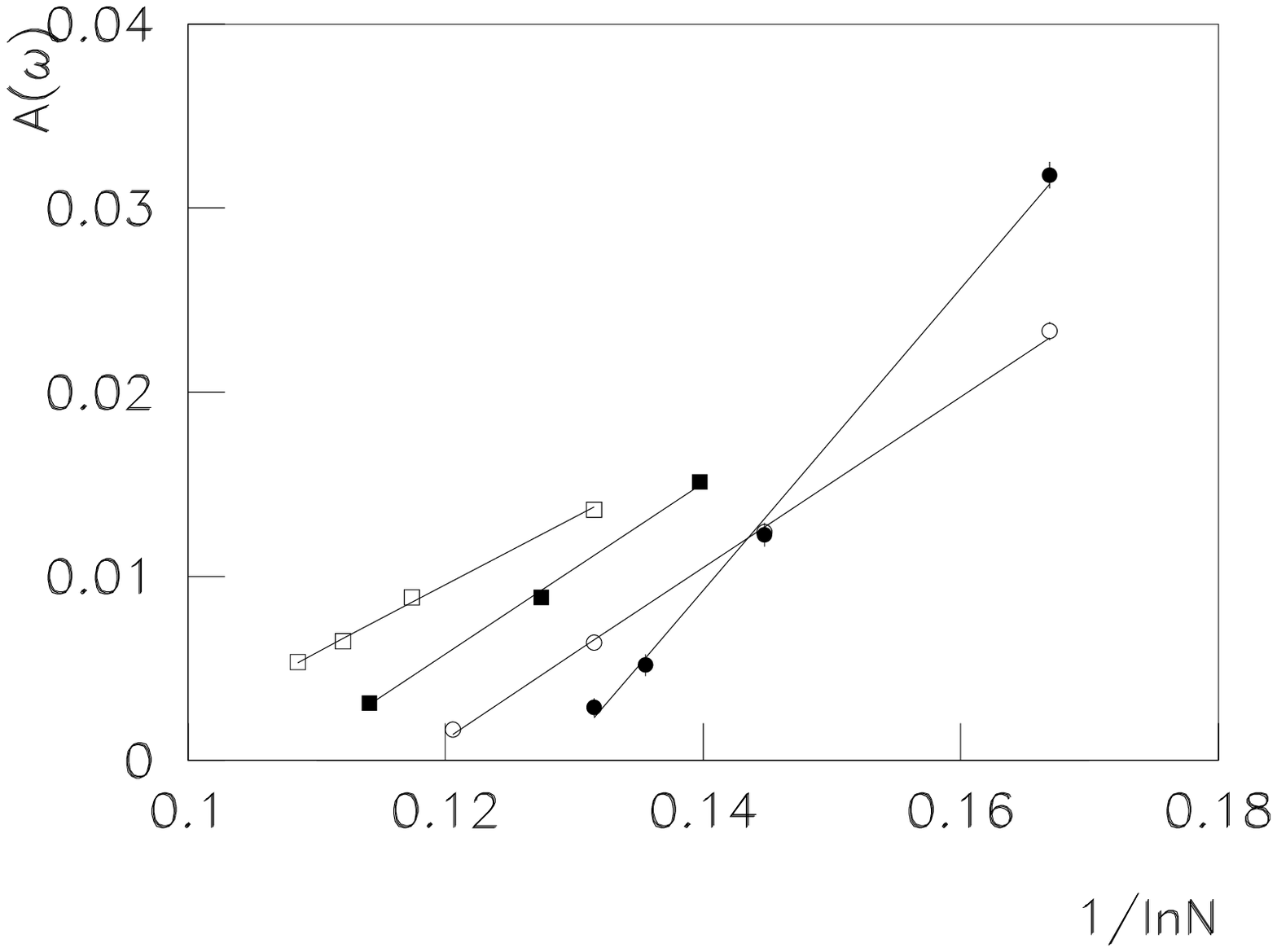}
Hysteresis loop areas $A(\omega)$ as functions of $1/\log N$ for the
15-state Potts model for $L=8$ (filled circles), 12 (filled circles),
16 (filled squares) and 20 (open squares). The lines show the best fits.
\endinsert

The direct measurements of $\tau(L)$ were obtained by constructing the
autocorrelation function and fitting its long-distance form to an
exponential. As a cross check, we measured local masses and looked for
plateaus as a signal that a single mass fit over a given range was
reasonable. The measurement procedure remains the same as in \ref{us}.
In all cases the fits were performed over a range which turned out to be
$\tau<t<4\tau$. The errors on $\tau(L)$ were, of course, reflections of
the errors on the autocorrelation function. These were obtained as the
dispersion between jack-knife blocks. We varied the number of jack-knife
blocks between 5 and 25. We took the lack of sensitivity of the means
and errors to the number of blocks as an indication that our error
estimates are reliable.

In order to test \eq{zinn} and measure \sod, we fitted the data on
$\tau(L)$ to the form
$$\log(\tau/L)\;=\;c L + c'.   \EQNO{fitsod}$$
Note that \sod{} is given by $c/2$. On the left hand side of
\eq{fitsod}, the division by $L$ takes care of the effects of
fluctuations around the instanton solution. It turns out that this term
in the fit is quite crucial. An attempt to perform the fit without this
factor was completely unsuccessful; $\chi^2$ values obtained increased by
almost an order of magnitude. In principle, one could perform a
three-parameter fit, leaving the power of $L$ in the pre-exponential
factor to be determined by the data. Unfortunately this requires more
lattice sizes than we had in this study. Our fits gave
$$2\sigma_{od}\;=\;
     0.263\pm0.009 \qquad\qquad(\chi^2=1.0/2 {\rm\ dof},\ 16\le L\le30).
     \EQNO{result2}$$
This result is compatible with perfect wetting.

\midinsert\vskip10truecm
\includegraphics{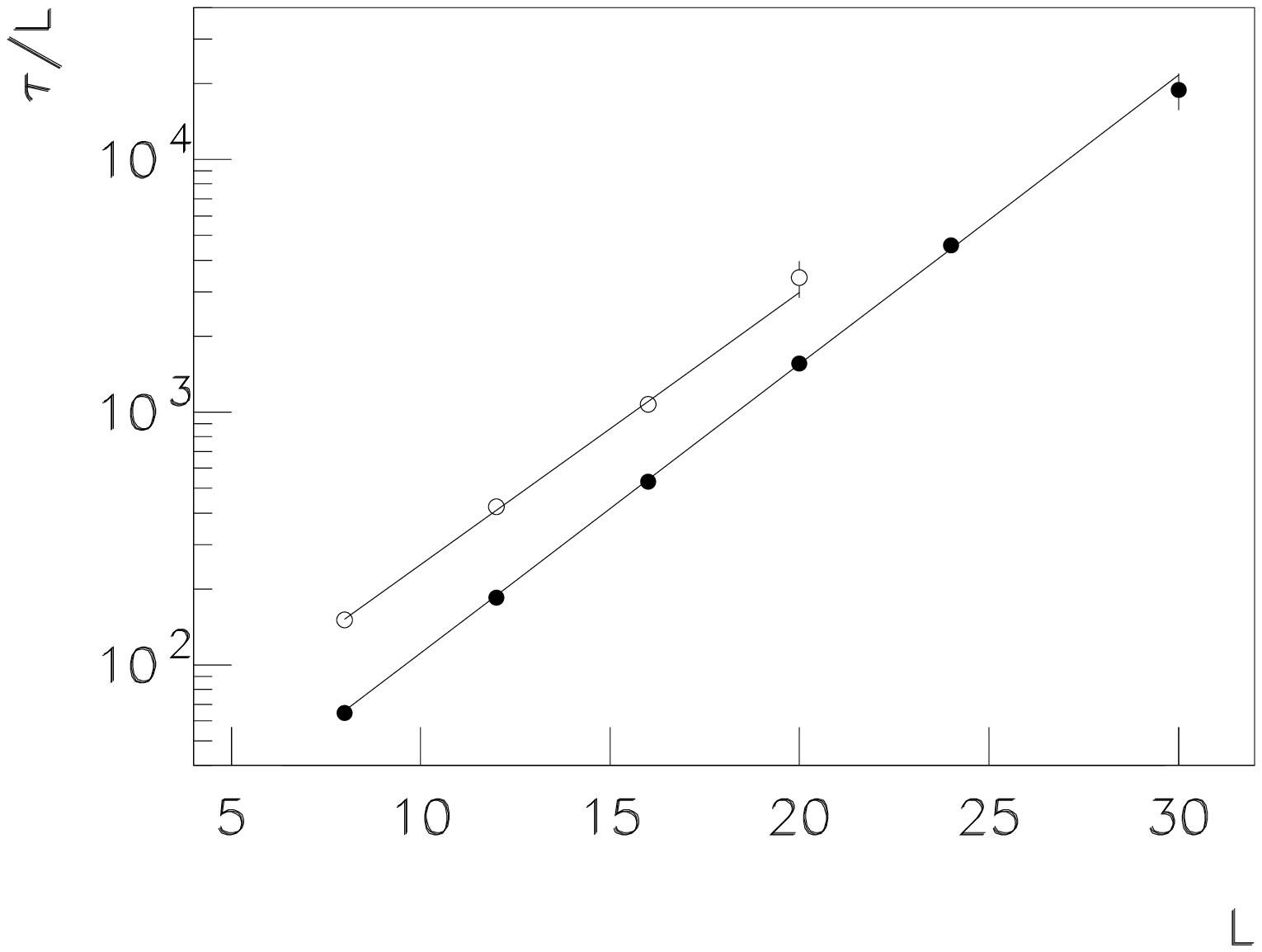}
\centerline{FIGURE \figtag{surften}}
We show the scaling of $1/\Omega$ (open circles) and directly measured
values of the autocorrelation time at $\beta_\T(L)$, $\tau$,
(filled circles) against the lattice size $L$.
The lines show the best fits of the form shown in \eq{fitline}.
The values of $1/\Omega$ have been multiplied by 2 for visibility.
\endinsert

Although the difference between the perfect wetting result for \sod{}
and our measurement is not statistically significant at the $3\sigma$
level, we believe it deserves comment. We find it difficult to regard
seriously the possibility that perfect wetting begins to break down
when $q$ drops to a number close to 15. More likely is that \eq{zinn}
has to be supplemented with a higher loop computation. It has been
argued \ref{zinn} that a loop-wise expansion of the pre-exponential
factor yields a power series in $L^{-d/2}$. It is a reasonable guess
that the two-loop term is marginally important for the
lattice sizes we have worked with. Then our observations would imply
that the coefficient of the term in $L^{-d}$ is positive. A computation
of this term would certainly be useful.

Finally we comment on previous numerical tests of \eqand{zinn}{deepak}.
A high-statistics study of the 10-state Potts model \ref{us} had
established that the autocorrelation time in equilibrium was determined
by the tunnelling phenomenon. However, this study had not been able to
observe even the dominant exponential behaviour in \eq{zinn}. It was
conjectured there that $L/\xi$ values used there were too small. The
recent work of \refs{bw}{anders} shows that this is indeed correct. In
that study the largest values of $L$ used were about $3\xi$, whereas this
study uses between $4\xi$ and $7\xi$.

Earlier studies of the scaling of
$A(\omega)$ with $\omega$ had parametrised the variation by power laws.
This is presumably correct for some systems, but the arguments of
\ref{deepak} must hold whenever the dominant dynamical mechanism is
nucleation and tunnelling. Magnetic hysteresis in the Ising model, or
the one-component $\phi^4$ theory should therefore be described by
\eq{deepak}. The contradictory results of \ref{others} were obtained
with values of $\omega$ much smaller than the ones we use. The
implication is that \eq{deepak} is not applicable to these.

We summarise the main results obtained in this study. The scaling law of
\eq{deepak} for the frequency dependence of hysteresis loop areas is
found to hold extremely well over three decades in frequency, and for a
variety of lattice sizes. This is strong evidence that in the particular
non-equilibrium situation at a first-order phase transition exemplified
by hysteresis, the dynamics is of nucleation. Furthermore, the
expression in \eq{zinn} is found to describe the finite-size scaling of
the autocorrelation times, showing that an instanton based description
of the equilibrium dynamics is valid. A proper treatment of fluctuations
around the instanton is observed to be crucial for the description of
the data. The dynamics then allows the extraction of the surface tension.
For the 15-state Potts model we find that perfect wetting holds. A
statistically insignificant discrepancy can be attributed to the neglect
of two-loop terms in the treatment of the fluctuation determinant. It
should be emphasised that this makes the present computation one of the
most accurate measurements of a surface tension to date.

\vfil\eject
\bigskip\centerline{\sectnfont REFERENCES}\bigskip
\item{\reftag{borgs})}
   C.~Borgs and R.~Koteck\'y, \JSP 61 (1990) 79;\lb
   C.~Borgs, R.~Koteck\'y and S.~Miracle-Sol\'e, \JSP 62 (1991) 529;\lb
   C.~Borgs and W.~Janke, \PRL 68 (1992) 1738.
\item{\reftag{fss})}
   S.~Gupta, A.~Irb\"ack and M.~Olsson, preprint HLRZ 23/93 and
   LU-TP-93-6, \NP B, in press.
\item{\reftag{billo})}
   A.~Billoire, R.~Lacaze and A.~Morel, \NP B370 (1992) 773;\lb
   B.~Berg, A.~Billoire and T.~Neuhaus, Saclay preprint SPhT-92/120.
\item{\reftag{sten})}
   K.~Binder, \ZP B 43 (1981) 119;\lb
   K.~Kajantie, L.~K\"arkkainen and K.~Rummukainen, \PL B223 (1989) 213;\lb
   J.~Potvin and C.~Rebbi, \PRL 62 (1989) 3062;\lb
   K.~Jansen \etal, \NP B322 (1989) 693.
\item{\reftag{zinn})}
   J.~C.~Niel and J.~Zinn-Justin, \NP B280 [FS18] (1987) 355.
\item{\reftag{deepak})}
   D.~Dhar and P.~B.~Thomas, preprint TIFR/TH/92-32.
\item{\reftag{bw})}
   E.~Buffenoir and S.~Wallon, Saclay preprint, SPhT/92-077.
\item{\reftag{bj})}
   C.~Borgs and W.~Janke, preprint HLRZ 54/92 and FUB-HEP 13/92.
\item{\reftag{us})}
   A.~Billoire, R.~Lacaze, A.~Morel, S.~Gupta, A.~Irba\"ck and B.~Petersson,
   \NP B358 (1991) 231.
\item{\reftag{matthias})}
   M.~Ohlsson, B.~Sc.\ Thesis, University of Lund, unpublished.
\item{\reftag{anders})}
   S.~Gupta and A.~Irb\"ack, \PL B286 (1992) 112.
\item{\reftag{others})}
   M.~Rao, H.~R.~Krishnamurthy and R.~Pandit, \PR B42 (1990) 856;\lb
   W.~S.~Lo and R.~A.~Pelcovits, \PR A42 (1990) 7471.
\vfil\end